\documentclass[]{spie}  
\usepackage[]{graphicx}
\usepackage[]{caption}
\usepackage[]{mathrsfs}

\newcommand{\bicep}{B{\sc icep}}

\title{Optimization and sensitivity of the Keck Array}

\author{S. Kernasovskiy,\supit{a,b}
P. A. R. Ade,\supit{c} 
R.W. Aikin,\supit{d}
M. Amiri,\supit{e} 
S. Benton,\supit{f} 
C. Bischoff,\supit{g} 
J.J. Bock,\supit{d,h}
J. A. Bonetti,\supit{h} 
J. A. Brevik,\supit{d} 
B. Burger,\supit{e} 
G. Davis,\supit{e} 
C.D. Dowell,\supit{d,h} 
L. Duband,\supit{i} 
J. P. Filippini,\supit{d} 
S. Fliescher,\supit{j}
S.R. Golwala,\supit{d} 
M. Halpern,\supit{e} 
M. Hasselfield,\supit{e} 
G. Hiltion,\supit{k} 
V.V. Hristov,\supit{d} 
K. Irwin,\supit{k} 
J. M. Kovac,\supit{g} 
C. L. Kuo,\supit{a,b}
E. Leitch,\supit{l} 
M. Lueker,\supit{d,h} 
C.B. Netterfield,\supit{f} 
H. T. Nguyen,\supit{d,h} 
R. O'Brient,\supit{d,h} 
R. W. Ogburn IV,\supit{a,b} 
C. L. Pryke,\supit{j} 
C. Reintsema,\supit{k}
J.E. Ruhl,\supit{m} 
M.C. Runyan,\supit{d} 
R. Schwarz,\supit{j} 
C. D. Sheehy,\supit{j} 
Z. Staniszewski,\supit{d,h} 
R. Sudiwala,\supit{c} 
G. Teply,\supit{d} 
J. E. Tolan,\supit{a,b} 
A. D. Turner,\supit{h} 
A. Vieregg,\supit{g}
D. V. Wiebe,\supit{e} 
P. Wilson,\supit{h} 
C. L. Wong\supit{g}
\skiplinehalf
\supit{a}Stanford University, 382 Via Pueblo Mall, Stanford, CA 94305, USA; \\
\supit{b}Kavli Institute for Particle Astrophysics and Cosmology (KIPAC), Sand Hill Road 2575, Menlo Park, CA 94025, USA; \\
\supit{c}Dept. of Physics and Astronomy, University of Wales, Cardiff, CF24 3YB, Wales, UK; \\
\supit{d}California Institute of Technology, 1200 E. California Blvd., Pasadena, CA 91125 USA; \\
\supit{e}Department of Physics and Astronomy, University of British Columbia, 6224 Agricultural Road, Vancouver, BC V6T1Z1, Canada; \\
\supit{f}Department of Physics, University of Toronto, Toronto, ON M5S 1A7, Canada; \\
\supit{g}Harvard-Smithsonian Center for Astrophysics, 60 Garden Street, Cambridge, MA 02138; \\
\supit{h}Jet Propulsion Laboratory, 4800 Oak Grove Dr., Pasadena, CA 91109, USA; \\
\supit{i}Service des Basses Tempratures, DRFMC, CEA-Grenoble, 17 rue des Martyrs, 38054 Grenoble Cedex 9, France; \\
\supit{j}School of Physics and Astronomy, University of Minnesota, 116 Church Street S.E.,Minneapolis, MN 55455; \\
\supit{k}NIST Quantum Devices Group, 325 Broadway, Boulder, CO 80305, USA; \\
\supit{l}University of Chicago, KICP, 933 E. 56th St., Chicago, IL 60637 USA; \\
\supit{m}Physics Department, Case Western Reserve University, Cleveland, OH 44106 USA; \\
}

\authorinfo{Send correspondence to S. Kernasovskiy, \\E-mail: sstokes@stanford.edu}

\begin{document}
  \maketitle

\begin{abstract}
The Keck Array (SPUD) began observing the cosmic microwave background's polarization in the winter of 2011 at the South Pole.  The Keck Array follows the success of the predecessor experiments \bicep\ and \bicep2\cite{Chiang:2009xsa}, using five on-axis refracting telescopes.  These have a combined imaging array of 2500 antenna-coupled TES bolometers read with a SQUID-based time domain multiplexing system.  We will discuss the detector noise and the optimization of the readout.  The achieved sensitivity of the Keck Array is 11.5 $\mu \mathrm{K}_{CMB} \sqrt{s}$ in the 2012 configuration.
\end{abstract}



\section{INTRODUCTION}
\label{sec:intro}

Inflation, the theory that the universe experienced exponential expansion in its first fraction of a second, was originally introduced as a way to solve the horizon problem--why the universe was nearly homogenous and geometrically flat\cite{1981PhRvD..23..347G}.  Since then, it predicted many phenomena since confirmed by observation: gaussianity, scale-invariance, and adiabaticity.\cite{Netterfield:2001yq,Komatsu:2008hk}.  It also is predicted to have produced a gravitational wave background.  The tensor perturbations generated would leave a signature in the curl component of the polarization (B-mode) of the Cosmic Microwave Background radiation (CMB)\cite{PhysRevLett.78.2058,0004-637X-482-1-6}.  The curl-free component of the polarization (E-mode) is dominated by the scalar, mass density perturbations.  The ratio of the tensor to scalar perturbations ($r$) depends on the energy scale of inflation.

The DASI experiment was the first to detect the polarization of the CMB\cite{Kovac:2002fg}, detecting the E-mode polarization.  The E-mode polarization power spectrum is determined by the same physics as the temperature power spectrum and thus provided a good test of the CMB paradigm.  The B-mode polarization signal is much smaller than the E-mode, and the current upper limits on $r<0.21$ (95\% CL) actually come from WMAP and SPT\cite{Keisler:2011aw} using the temperature information.  BICEP provides the best upper limit from the B-mode power spectrum at $r<0.72$ (95\% CL)\cite{Chiang:2009xsa}.

The Keck Array (aka SPUD) is a ground based polarimeter currently observing the CMB.  The Keck Array is the latest of the \bicep/\bicep2 family of experiments.  All three experiments use similar optical designs, \bicep2 uses a focal plane with the same detectors as the Keck Array, and the Keck Array transitioned to a closed-cycle He cooler and increased the number of receivers.  These experiments were designed specifically with the goal of measuring the imprint of gravitational waves from inflation on the polarization of the CMB.  Measuring such a signature will require a lot of sensitivity in large angular scale ($\ell \sim 100$) polarization.  Our strategy is to integrate deeply on a 800 $\mathrm{deg}^2$ patch of sky with a $0.5\ \mathrm{deg}$ FWHM beam. 

This paper is focused on the Keck Array and its sensitivity.  Several companion papers presented at this conference focus on the current status of \bicep2 and the Keck Array (Ogburn et al.\cite{2012SPIE.Ogburn}), the Keck Array optical performance (Vieregg et al.\cite{2012SPIE.Vieregg}), the performance of the dual planar antennas (O'Brient et al.\cite{2012SPIE.Obrient}), and the thermal stability performance of \bicep2 (Kaufman et al.\cite{2012SPIE.Kaufman}).

\section{Instrument}
\label{sec:instrument}

The Keck Array consists of a set of five telescopes, each of which is very similar in design to \bicep2\cite{Sheehy:2011yf,2010SPIE.Aikin,2010SPIE.Ogburn}.  Each telescope consists of a 26.4 cm aperture, refractive optics focusing light onto a focal plane of 256 dual-polarization detector pairs.  Currently, all five telescopes are observing at 150 GHz.  In the future, we have the ability to switch the optics and detectors of individual telescopes to either a frequency of 100 GHz or 220 GHz.  This modular design allows for tight systematic control, including a complete shielding from the ground, stable temperature of the optics and loading on the focal plane, and end-to-end optical measurements.

The receiver and all of the optics, are cooled to 4K using a closed-cycle He pulse tube system.  This is the main difference in design between the Keck Array and \bicep2, which uses liquid helium.  The focal plane is cooled to 270 mK using a 3He/4He 3-stage sorption fridge.  

The first 3 receivers were deployed to the South Pole in the Martin A Pomerantz Observatory (MAPO) for the 2011 season, using the 3-axis mount originally built for the DASI experiment. The remaining 2 receiving were installed and observations with the full 5 receivers commenced in 2012.

\subsection{Detectors}
\label{sec:detectors}

The detectors, developed at JPL for joint use in \bicep2, SPIDER and the Keck Array experiments, consists of phased array antennas feeding into Ti transition-edge sensor (TES) bolometers,\cite{Kuo:2009rj,Orlando:2010zfa} described in detail in O'Brient et al.\cite{2012SPIE.Obrient}  These detectors are lithographically patterned, making it easier to produce many detectors at a time.  Each camera element consists of two phased antenna arrays, with orthogonal polarization direction.  Each signal is then bandpass filtered and terminated into a Ti TES suspended in a SiN membrane. A single 4" Si tile contains 64 detector pairs and a focal plane unit has 4 tiles.  In each focal plane, 16 detectors are left dark.  Dark detectors consist of the complete TES island structure, but are not connected to their corresponding antennas.

The detectors in the Keck Array are the result of extensive fabrication and testing.  The earliest of the currently used detectors were made in the summer of 2010, and the latest in late 2011.  The thermal connection between the TES and the bath can be modeled using load curves by sweeping the detector bias and reading the output current with the focal plane held at different temperatures.  For small temperature changes $\delta T$ around the transition temperature $T_c$, the load power to the bath $P_{bath}$ is approximated by Equation \ref{eqn:pbath}.

\begin{equation}
P_{bath}=K(T^{\beta+1}-T_{bath}^{\beta+1})\approx P_{bath_0}+G_c\delta T
\label{eqn:pbath}
\end{equation}
The Keck Array detectors have thermal conductivities $G_{c}$ of $40-90$ pW/K, $T_{c}$ of $470-530$ mK and thermal conductance exponent $\beta$ of $2.5$.  The $G_{c}$ and $T_{c}$ were designed to minimize the detector noise, while leaving a margin of safety to ensure the detectors are not saturated during observations.  Under standard observing conditions at the South Pole, the focal planes are cooled to $270$ mK, and the Joule power $P_J$ is 5-10 pW in transition.  The optical loading ($P$), measured by comparing the saturation power of the dark detectors to the light detectors is $2-4$ pW.  The detector parameters are summarized in Table \ref{tab:detsummary}.

\begin{table}[h]
\begin{center}
\begin{tabular}{|l||l|l|l|l|l|l|l|l|l|l|l|l|l|l|l|l|l|l|l|l|}
\hline
\rule[-1ex]{0pt}{3.5ex}  Receiver & Rx0 & Rx1 & Rx2 & Rx3 & Rx4\\
\hline
\rule[-1ex]{0pt}{3.5ex}  $G_c (pW/K)$ & 80 & 50 & 80 & 40 & 60\\
\hline
\rule[-1ex]{0pt}{3.5ex}  $T_c (mK)$ & 530 & 500 & 530 & 470 & 500\\
\hline
\rule[-1ex]{0pt}{3.5ex}  $P_{J} (pW)$ & 7.7 & 6.7 & 7.4 & 4.9 & 7.2\\
\hline
\end{tabular}
\caption{Summary of detector parameters for the Keck Array receivers.  There is significant variation within receivers by detector tile, but the values are averaged here for conciseness.}
\label{tab:detsummary}
\end{center}
\end{table}

The optical efficiencies of the detectors are measured in the lab as part of the standard testing regimen.  The measurements are taken by placing a beam-filling, microwave-absorbing cone over the vacuum window of the receiver at room temperature ($\sim 295$K) and at liquid nitrogen temperature ($\sim 74$K at the South Pole).  The optical efficiency $\eta$ is defined as the fraction of power that the bolometer measures compared to total light power input.
\begin{equation}
P = \eta \int_{\nu_1}^{\nu_2} \frac{h \nu}{\exp(h\nu/kT)-1}\,\mathrm{d}\nu
\label{eq:opeff}
\end{equation}
where $P$ is the measured power with the bolometer, and $\nu_1$ and $\nu_2$ are the band defining frequencies.  At 150 GHz, both of these temperatures are well in the Rayleigh-Jeans limit ($h\nu\ll kT$).  Equation \ref{eq:opeff} reduces to $\Delta P = \eta k_B \nu \frac{\Delta \nu}{\nu} \Delta T$, where $\frac{\Delta \nu}{\nu}$ is the fractional spectral bandwidth (0.22), $\nu$ is the band center frequency (145 GHz), and $\Delta T$ is the difference in temperature of the loads.  From these measurements, the median optical efficiency $\eta$ is $\sim 30\%$.

\subsection{Readout}
\label{sec:readout}

We apply a voltage bias to the TESs and read out the current from the detectors using time-division multiplexed superconducting quantum interference devices (SQUID)\cite{deKorte}.  These were developed at NIST and are used in numerous other CMB/submillimeter experiments.  This multiplexing system consists of 3 stages of SQUIDs.  Each detector is inductively coupled to a first stage SQUID (SQ1).  A set of 33 SQ1s are inductively-coupled into a summing coil leading to a single second-stage SQUID (SQ2).  They are multiplexed by simply turning the SQ1 biases on and off.  When its bias is turned off, the SQ1 is superconducting and does not contribute to the summing coil.  The output of the SQ2 then leads to a high gain SQUID series array (SSA) before exiting the cryostat.  The SQ1s and SQ2s are placed on the focal plane unit along with the detectors at 270 mK.  The SSAs are cooled to 4K.  This readout system is also described by Ogburn et al.\cite{2010SPIE.Ogburn,2012SPIE.Ogburn}.

The electronics for the SQUID multiplexers (Multichannel Electronics or MCE) were developed at UBC\cite{Battistelli2008}.  These electronics control the current bias and flux feedback (to center the SQUID $V(\phi)$ modulation curve) for every SQUID.  They also bias the TESs, with only 1 common bias line used per 32 TESs.  The detectors sharing a single bias line are all drawn from the same detector tile and have similar enough properties that this is acceptable.

\section{Observations}
\label{sec:observations}

The Keck Array observes the CMB in a patch of 800 $\mathrm{deg}^2$ in the southern hemisphere known as the ``Southern Hole''.  This patch of sky has very low dust and synchrotron foreground emissions.  It overlaps with the observing regions of several other experiments located in the southern hemisphere including BICEP/\bicep2 and SPTPol.  150 GHz is near the minimum of expected astronomical foreground contaminations for this patch.

The observing strategy is to scan in azimuth while stepping in elevation.  Every 2 days, the entire telescope, ground shielding baffles, optics and detectors are rotated about the boresight.  We observe at 4 different boresight rotations: 2 sets of 180$^\circ$ rotations offset 45$^\circ$ from each other, completing the Stokes Q and U maps.  The 180$^\circ$ rotation takes advantage of the small size of the Keck Array to suppress systematics.  In every 48 hour cycle, the Keck Array observes 4$\times$9 hours on the CMB patch, 1$\times$6 hours on the galactic plane, and spends 6 hours on cryogen servicing.

\subsection{Calibrations}
\label{sec:calib}

During the Austral summer we perform instrument calibrations.  A source is placed a distance of 200m away from the Keck Array and is observed using a large, 45 $\mathrm{deg}$, flat mirror to produce maps of the detector beams.  The beams have a median FWHM of $0.5 \mathrm{deg}$.  The optical characteristics of the Keck Array will be more fully described by Vieregg et al. in these proceedings.

During observations, calibration measurements are routinely taken.  Every hour, the Keck Array does a 1 $\mathrm{deg}$ elevation nod on the sky (elnod) in order to calibrate the relative gain between each polarization pair of detectors.  The TES biases are also swept to measure the operating resistance and the saturation power.  Every 10 hours, we do a larger dip on the sky and a full load curve.


\section{Detector Noise}
\label{sec:noise}

   \begin{figure}
   \begin{center}
   \begin{tabular}{c}
   \includegraphics[height=6cm]{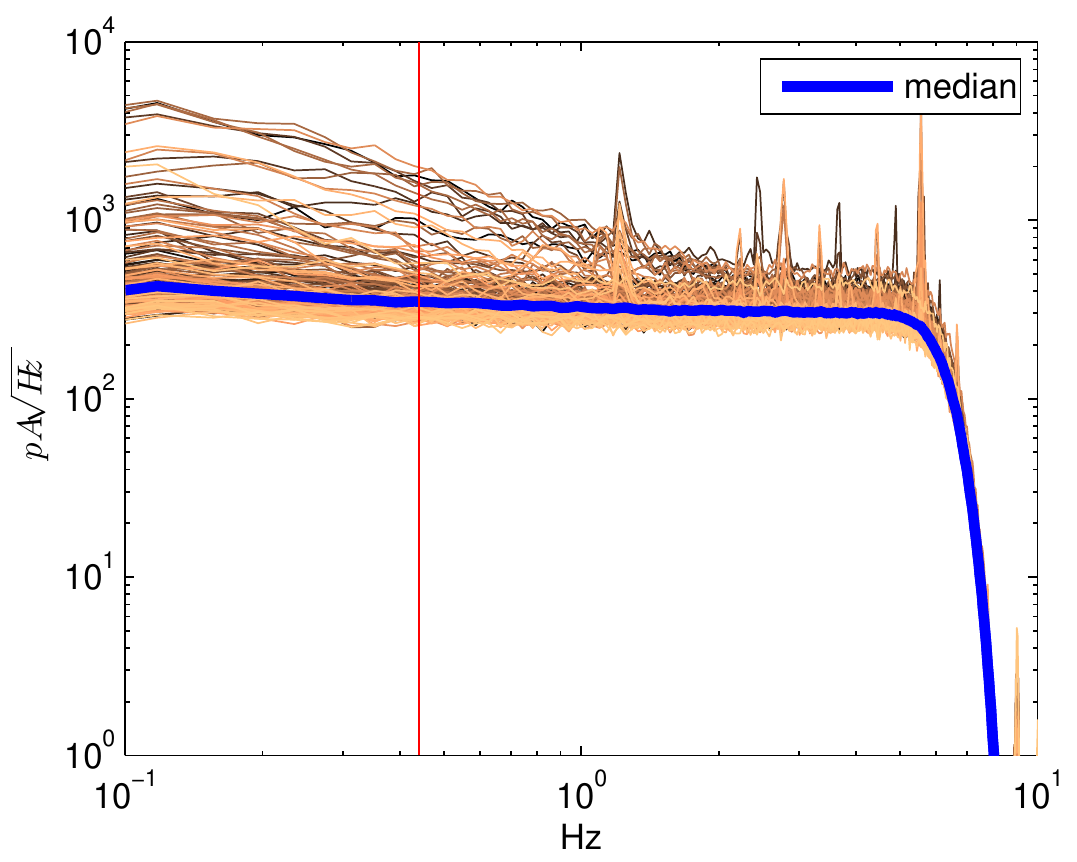}
   \end{tabular}
   \end{center}
   \caption[example]
   { \label{fig:pairdiffnoise}
  Polarization pair differenced noise spectrum for all of the pairs in one receiver.  Most of the 1/f noise induced by the atmosphere is common mode and is removed by differencing the pairs.  The red line indicates $\ell$=100 when scanning at 1.6 $\mathrm{deg}/\mathrm{s}$ in azimuth.}
   \end{figure}

We measure the noise of the detectors in several ways.  In normal data taking mode, we multiplex, filter, and downsample the timestream to 20 Hz.  The targeted range of CMB modes is $\ell\sim 30-300$, with a peak from primordial gravitational B-modes expected at $\ell\sim 80$.  In combination with a scan velocity of 1.6 deg/s in azimuth, the B-mode signal will appear at frequencies between 0.1 and 1 Hz in the recorded timestreams.  Thus saving the data at 20 Hz is more than sufficient to retain the information on the targeted $\ell$ range.  This $\ell$ range is also sufficiently high to avoid 1/f induced by variations of the atmosphere into the polarization pair-differenced data.  Figure \ref{fig:pairdiffnoise} shows the noise spectrum in pair-differenced observational data.

   \begin{figure}
   \begin{center}
   \begin{tabular}{c}
   \includegraphics[height=5.5cm]{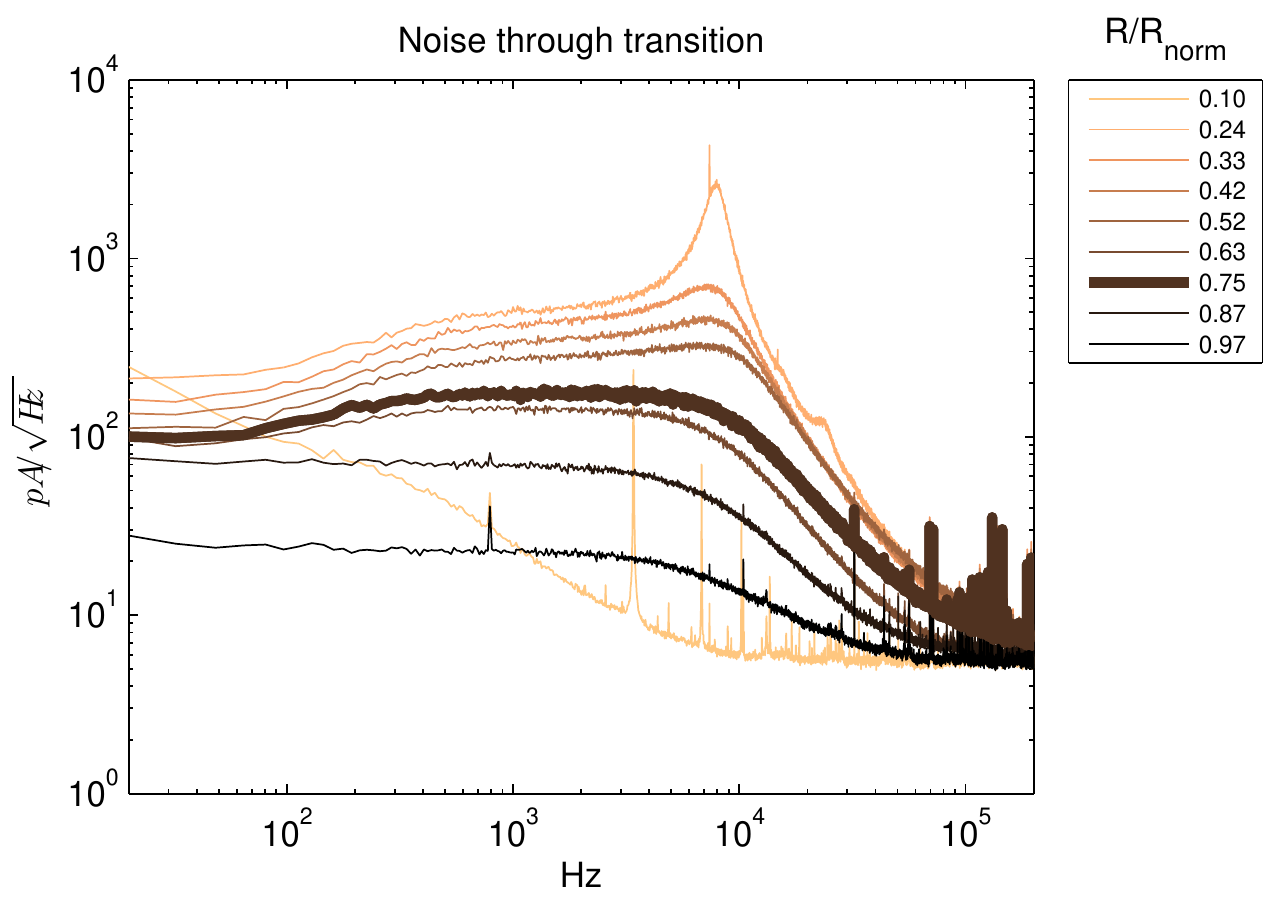}
   \includegraphics[height=5.5cm]{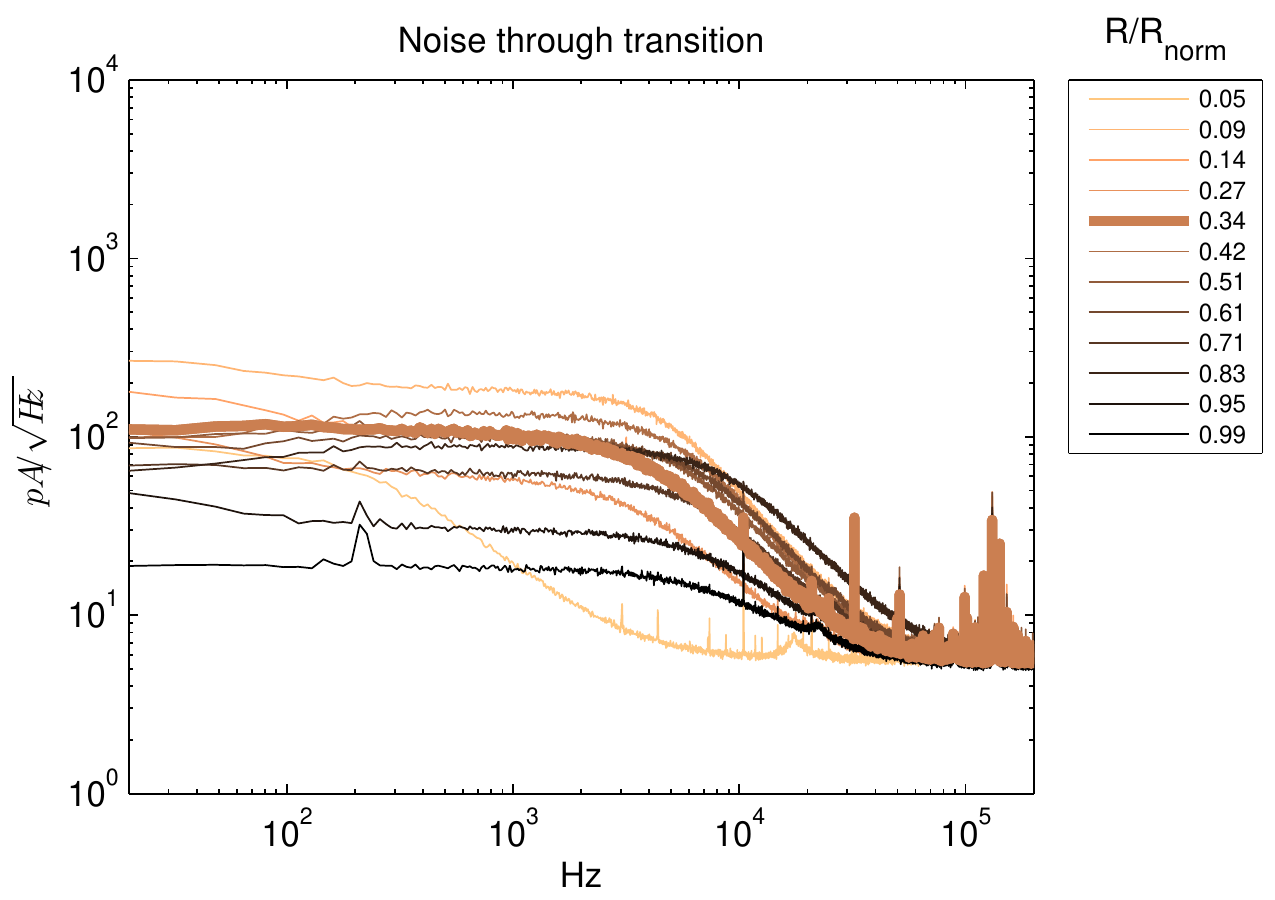}
   \end{tabular}
   \end{center}
   \caption[example]
   { \label{fig:noisevbias}
  High frequency noise spectra for a single detector plotted for different fractional TES resistances $R/R_{normal}$.  The detectors are normal conducting at high biases (black) and superconducting at low biases (yellow).  The bolded noise spectra are at the biases used during standard observations.  The detector on the left is a detector with substantial excess noise, which shows up as increased noise between $10^2$ and $10^4$ Hz.  The detector on the right is typical of those with less excess noise at low biases.}
   \end{figure}

We are able to record data from single detectors at higher sampling frequencies by avoiding the multiplexing step in the readout described in Section \ref{sec:readout}. An example of single detector noise spectra, taken at 400kHz, as it is biased through transition is in Figure \ref{fig:noisevbias}.  We have taken such noise spectra from every detector as they are biased through transition to characterize their noise properties.

\begin{figure}
   \begin{center}
   \begin{tabular}{c}
   \includegraphics[height=6cm]{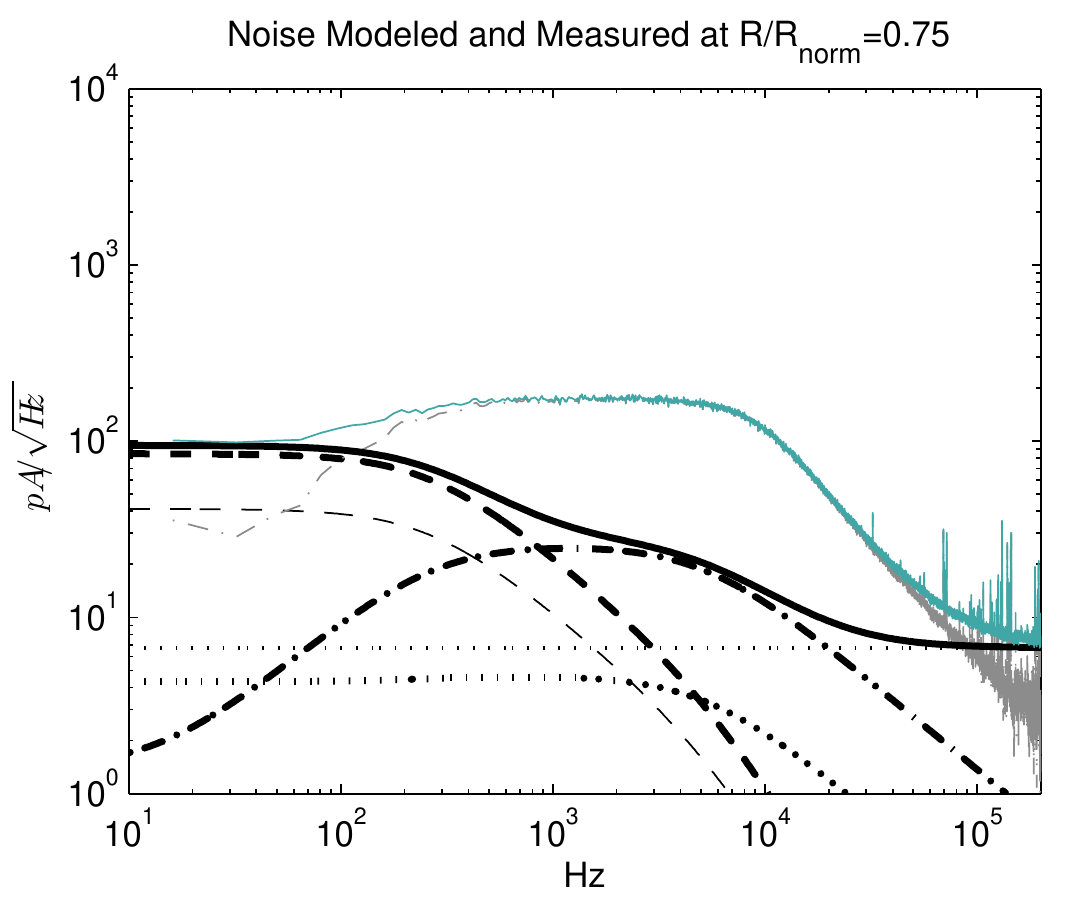}
   \includegraphics[height=6cm]{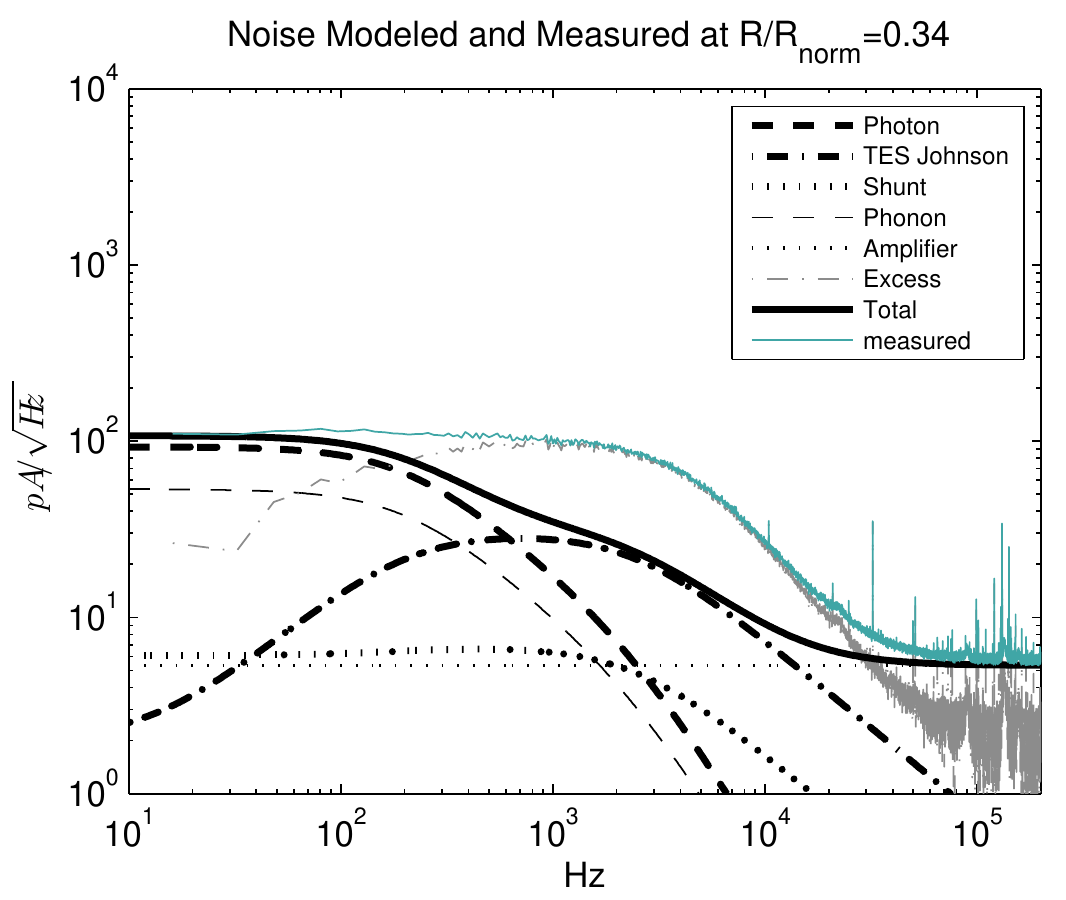}
   \end{tabular}
   \end{center}
   \caption[example]
   { \label{fig:noisemodel}
  High frequency noise spectra and noise model for a single $R/R_{normal}$.  These are the same detectors as in Figure \ref{fig:noisevbias}: the detector on the left has excess noise as low biases, while the detector on the right has less.  The excess noise is marked as the measured spectra minus the model.  The TES bias plotted is what is used during normal observations.  This is an example at $R=0.75\times R_{normal}$ and $R=0.35\times R_{normal}$.}
   \end{figure}

In the 2010 SPIE proceedings, Brevik et al.\cite{2010SPIE.Brevik} presented a detailed model of the noise of \bicep2 detectors.  Since the detectors of \bicep2 and the Keck Array follow the same design, they have a similar noise profile.  The noise spectrum is composed of photon noise, phonon noise, Johnson noise, amplifier noise, and an excess noise which seems to peak broadly at ~1kHz.  Figure \ref{fig:noisemodel} shows this model for the same detectors as Figure \ref{fig:noisevbias}, biased at the standard operation configuration.  

The excess noise's spectral signature is similar to Johnson noise.  This is not a new phenomenon -- it has been observed in other TESs.\cite{IrwinHilton,CabreraTES}  Several articles discuss models of this noise\cite{Irwin2006718,Galeazzi_2011}, noting that it is an expected fundamental noise from a TES.  Although the noise is not modeled in this paper, it is worth noting that the detectors do not all have the same amount of excess noise at low biases.  Figure \ref{fig:noisevbias} shows a detector with little excess noise at low biases and a detector with substantial excess at low biases.  Either way, 100Hz is above the frequency band of interest, and the excess noise should not increase the overall noise levels as long as it is not aliased into the science band from multiplexing.  This will be discussed more in Section \ref{sec:mux}.

The photon noise is dominant at low frequencies and is modeled by Equation \ref{eq:photon}.  The noise-equivalent power (NEP) is:
\begin{equation}
\mathrm{NEP}^2_{\mathrm{photon}}=2h\nu Q_{\mathrm{load}}+\frac{2Q_{\mathrm{load}}}{\nu \frac{\Delta\nu}{\nu}}
\label{eq:photon}
\end{equation}
where $\nu$ is the frequency, $\frac{\Delta\nu}{\nu}$ is the spectral bandwidth.  The photon loading $Q_{\mathrm{load}}$ can be estimated by the temperatures of the optics and the sky emissivity.  The photon noise in the Figure \ref{fig:noisemodel} is consistent with a loading of 4.5 pW, or 25 Kelvin (Rayleigh-Jeans equivalent) at zenith.  This is consistent with what we expect for a warm austral summer day at the South Pole.  During the winter observing season, the photon loading is closer to 3 pW for these detectors.  

The phonon noise also contributes significantly at low frequencies.  These are thermodynamic fluctuations associated with the thermal conductance.\cite{IrwinHilton}.  This is modeled in Equation \ref{eq:phonon}.
\begin{equation}
\mathrm{NEP}^2_{\mathrm{TFN}}=4k_B T_c^2 G F(T_c,T_{bath})
\label{eq:phonon}
\end{equation}
where $F(T_c,T_{bath})$ accounts for non-linear thermal conductance and is estimated to be $\sim 0.5$ for our configuration.\cite{2010SPIE.Brevik,Mather:82}.  The measured thermal conductances $G_c$ and transition temperature $T_c$ from Table \ref{tab:detsummary} were inputs for the phonon noise.

The Johnson noise, or thermodynamic fluctuations of the electrical resistance, is estimated by Equation \ref{eq:johnson} at DC:

\begin{equation}
\mathrm{NEP}^2_{\mathrm{Johnson}}=\mathrm{NEP}^2_{\mathrm{TES}}+\mathrm{NEP}^2_{\mathrm{shunt}}=4k_b T_c R_{\mathrm{TES}} I_{\mathrm{TES}}^2 \frac{1}{\mathscr{L}^2}+ 4 k_b T_{sh} R_{sh} I_{\mathrm{TES}}^2 \frac{(\mathscr{L}-1)^2}{\mathscr{L}^2}
\label{eq:johnson}
\end{equation}
The Johnson noise is suppressed by the TES thermal feedback loop gain $\mathscr{L}$, making the Johnson noise a sub-dominate component at low frequencies.  The TES loop gain $\mathscr{L}=\frac{P\alpha}{GT}$ is dependent on the steepness of the transition $\alpha$, which is estimated to be about 100 based on temperature versus resistance data.  The current $I_{\mathrm{TES}}$ and resistance $R_{\mathrm{TES}}$ of the TES were measured using a load curve for calibration.  During observations, the current is roughly 10 $\mu$A, the TES resistance is 0.3-0.9 $R_{normal}$, the shunt resistors are 3 m$\Omega$, and the normal resistances of the TESs are about 60 m$\Omega$.   

The contributions from each of the noise terms are given in Table \ref{tab:detnoisesummary}.  These are the median fits across all of the good detectors within in a receiver. This noise was taken in April, 2012, when the optical loading was 2-3.5 pW.  The receivers with lower $G_c$ have less phonon noise, and the receivers with lower optical efficiency have less photon noise.

\begin{table}[h]
\begin{center}
\begin{tabular}{|l||l|l|l|l|l|l|l|l|l|l|l|l|l|l|l|l|l|l|l|l|}
\hline
\rule[-1ex]{0pt}{3.5ex}  Receiver & Rx0 & Rx1 & Rx2 & Rx3 & Rx4\\
\hline
\rule[-1ex]{0pt}{3.5ex}  Photon Noise (aW/$\sqrt{Hz}$) & 33 & 32 & 34 & 27 & 20\\
\hline
\rule[-1ex]{0pt}{3.5ex}  Phonon Noise (aW/$\sqrt{Hz}$) & 24 & 20 & 24 & 14 & 16\\
\hline
\rule[-1ex]{0pt}{3.5ex}  Johnson Noise (aW/$\sqrt{Hz}$) & 0.8 & 0.5 & 0.8 & 0.4 & 0.6\\
\hline
\rule[-1ex]{0pt}{3.5ex}  Amplifier Noise (aW/$\sqrt{Hz}$) & 2.0 & 2.0 & 2.4 & 2.0 & 2.5\\
\hline
\rule[-1ex]{0pt}{3.5ex}  Total Noise (aW/$\sqrt{Hz}$) & 41 & 37 & 41 & 32 & 25\\
\hline
\rule[-1ex]{0pt}{3.5ex}  Total Noise Aliased at 15 kHz (aW/$\sqrt{Hz}$) & 55 & 44 & 63 & 45 & 34\\
\hline
\rule[-1ex]{0pt}{3.5ex}  Total Noise Aliased at 25 kHz (aW/$\sqrt{Hz}$) & 46 & 39 & 48 & 36 & 29\\
\hline
\end{tabular}
\caption{Summary Table of detector noise contribution for the Keck Array receivers.  Again, there is significant variation within receivers by detector tile, but the values are averaged here for conciseness.}
\label{tab:detnoisesummary}
\end{center}
\end{table}

\section{Optimization}
\label{sec:opt}

Since 2011, we have put a significant amount of effort into improving the sensitivity of the system.  The single largest, and most obvious, was to add two more telescopes.  Two of the existing telescopes were upgraded, including replacing one of the focal planes.  We also spent time improving the SQUID feedback servo and increasing the multiplexing speed.

\subsection{TES Bias Selection}
\label{sec:tesbias}

\begin{figure}
   \begin{center}
   \begin{tabular}{c}
   \includegraphics[height=7cm]{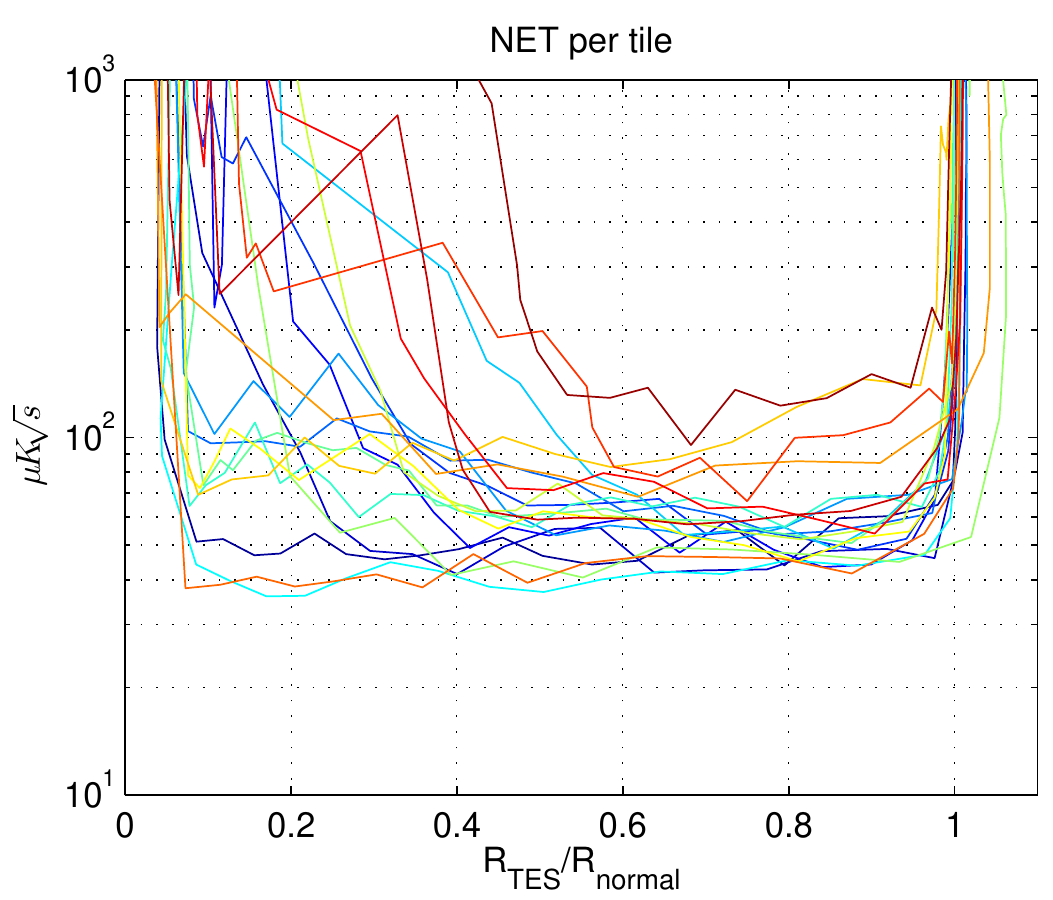}
   \end{tabular}
   \end{center}
   \caption[example]
   { \label{fig:NET}
The NET per detector tile as a function of TES fractional resistance.  Each line is a different tile for all 20 tiles.  There is variation in the shape of the NET versus resistance curve.  The detectors with little excess noise at low biases are flat bottomed, while those with significant amount rise in NET at low resistances.}
   \end{figure}

We selected the TES biases by sweeping through a range of bias currents, taking elevation nods on the sky to measure the optical response, followed by five minutes of noise data taken with the telescope stationary.  Figure \ref{fig:NET} shows the noise-equivalent temperature (NET) versus bias for each detector tile.

The TES bias is also constrained on the lower end by when the TES goes unstable.  At low biases, the thermal time constant and the electrical time constants can interfere and cause the TES to go into oscillations\cite{IrwinHilton}.  The NET per detector column (common MCE bias line) may still go down because some detectors are very sensitive in that regime.  However, the other detectors are unstable and can cause cross talk and other problems within the system.

\subsection{Multiplexing}
\label{sec:mux}

The noise performance was also improved by increasing the multiplexing speed.  We must multiplex quickly enough for the Nyquist frequency to exceed our noise bandwidth to avoid a large noise aliasing penalty.  In 2011, we multiplexed at 15 kHz.  We estimated that the noise was degraded by 15-20$\%$ by multiplexing at 15 kHz versus 25 kHz.  \bicep2 has shown a similar increase in sensitivity by moving to 25 kHz\cite{2010SPIE.Brevik,2012SPIE.Ogburn}.  Learning from \bicep2's experience, the Keck Array was able to deploy Nyquist chips with 2 $\mu$H bandwidth limiting inductors on four out of five focal planes, but the noise was still decreased by multiplexing faster.  The limit on multiplexing speed comes from the time needed for the multiplexing circuit to settle to its new voltage after each row switch.  

\begin{figure}
   \begin{center}
   \begin{tabular}{c}
   \includegraphics[height=6cm]{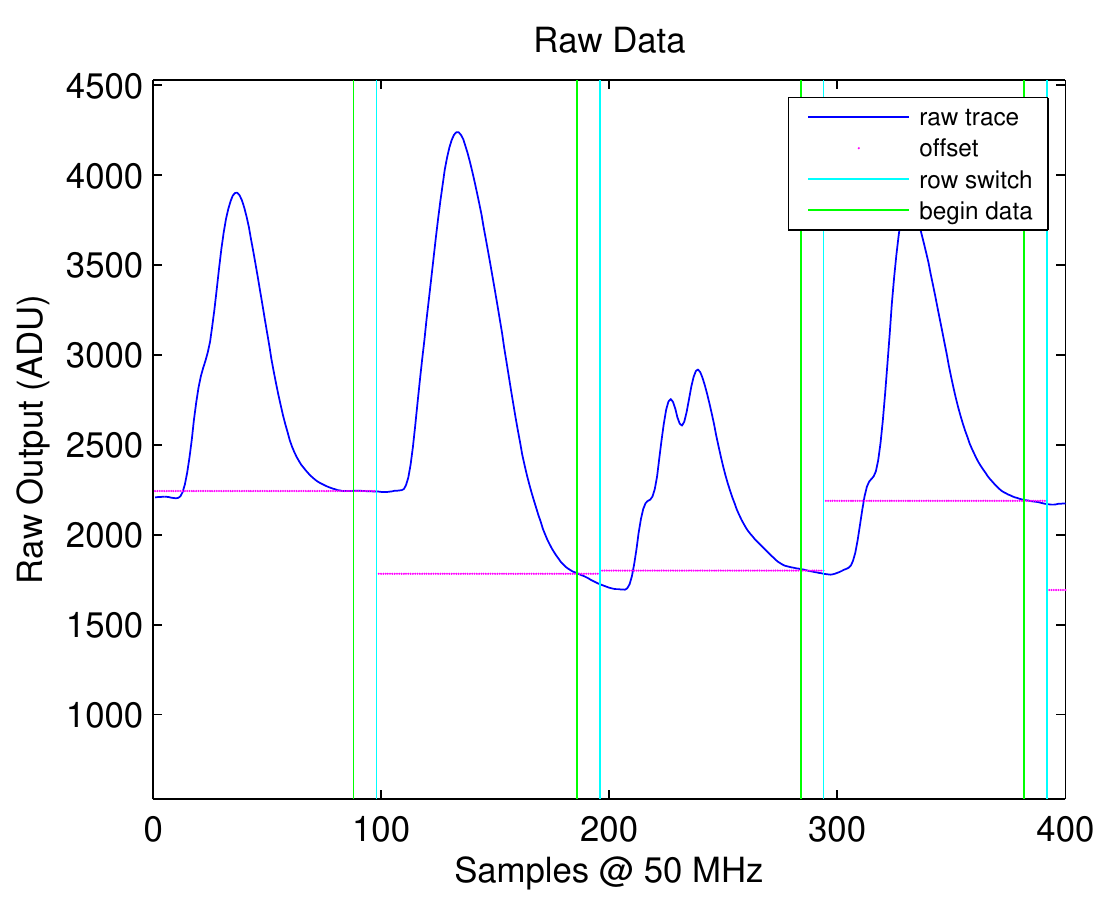}
   \includegraphics[height=6cm]{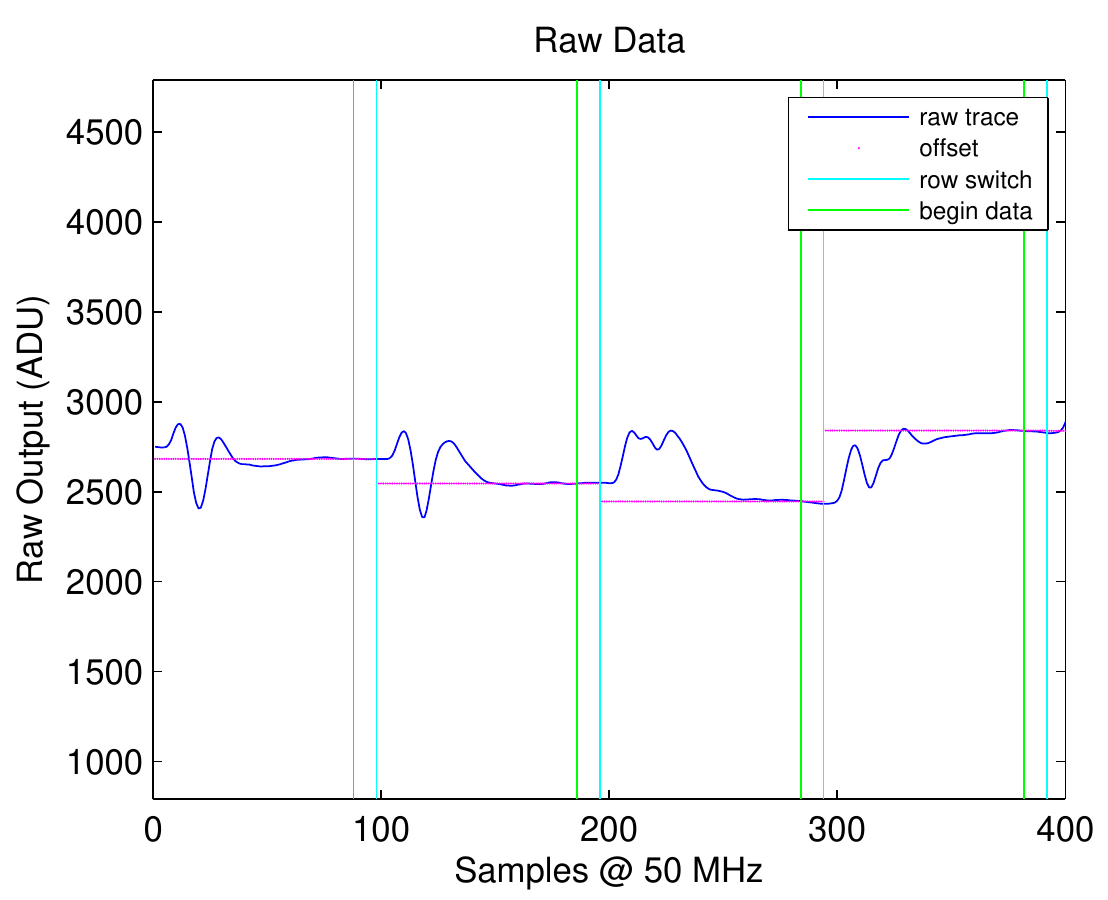}
   \end{tabular}
   \end{center}
   \caption[example]
   { \label{fig:raw}
Raw multiplexing time streams.  On the left is in the 2011 configuration.  On the right is 2012.  The blue line is the raw data trace.  At each cyan line, the SQ1 row is switched, and it must settle before data is taken at the green line.  The magenta line is the settling point for each detector.  The changes of decreasing the switching time of SQ1 feedback and maximizing the SQ2 dynamical resistance made it so the Keck Array could multiplex at 25 kHz, which decreased the aliased noise contribution.}
   \end{figure}

The time-domain SQUIDs switch between 33 rows.  At each row, it sets the SQ1 bias, the SQ1 feedback, and the SQ2 feedback (to appropriately center for each SQ1).  This has a settling time.  In 2011, the detectors and SQUIDs were allowed to settle for 88 clock cycles before 10 samples were added to form a data point.  With a 50MHz clock, this means each detector was recorded with a frequency of $f=50\mathrm{MHz}/(33 \times 98)=15\mathrm{kHz}$.  For Keck Array, the multiplexing frequency was limited by the settling time of the SQUIDs. This is evident in Figure \ref{fig:raw}, which is a raw data timestream as it switches between 4 detectors.

The settling times were improved significantly by decreasing the time spent applying the SQ1 feedback and maximizing the dynamical resistance ($R_{dyn} =\partial V/\partial I$) of the SQ2.  The first improvement works because there is a rise time to the switching transients.  The amplitudes of the transients were reduced significantly when the switching time was decreased.  The SQ1 feedback switching time had previously been increased in order to allow the MCE to calculate flux-jumps of the SQ1.  The second improvement is to minimize the settling time of the system.  The settling times are dominated by the SQUID2 to series array time constant of $\tau = L/R_{dyn}$, where $L$ is the inductance in that line.  The maximal $R_{dyn}$ turned out to be at a bias of $1.2$ times the critical current $I_{c-max}$ of the SQ2.  The critical current $I_{c-max}$ corresponds to where the SQ2 is no longer superconducting, has the largest amplitude $V(\phi)$ modulation curve and highest gain and therefore was previously the nominal biasing point.  These two changes resulted in much smaller transients and quicker settling times in the 2012 configuration, as seen in Figure \ref{fig:raw}.

\subsection{SQUID Feedback Servo Changes}
\label{sec:servo}

In normal data taking modes, the MCE servos on the first-stage SQUID.  This is because SQUIDs have optimal gain and are the most linear in the middle of their $\mathrm{V}(\phi)$ modulation curve.  In 2011, the Keck Array used a basic I-term servo, as is common with the MCE, and as was done with ACT and SCUBA-2\cite{Battistelli2008}.  This servo was simply $f(n) = e(n)+f(n-1)$, where $f$ is the applied SQ1 feedback at point $n$ and $e$ is the error signal.  The new time-constant limited I-term servo the Keck Array uses is $f(n) = e(n) + b\times f(n-1)$, where $b$ is a decay term less than unity.  With the basic I-term, the feedback signal can accumulate indefinitely, eventually overflowing.   This generates large amplitude currents in the multiplexer which can interfere with neighboring channels, inducing crosstalk as large as 0.1 $K_{CMB}$.   The time-constant limited I-term effectively discards old error data from the overall sum, keeping this term bounded, while still preserving servo linearity.

\section{Initial Sensitivity Estimates}
\label{sec:sensitivity}

The Keck Array has been observing with a full set of receivers and detectors since February, 2012.  The initial estimate of the sensitivity of the array is 11.5 $\mu \mathrm{K}_{CMB} \sqrt{s}$ in the 2012 configuration.

\subsection{Map based estimates}
\label{sec:map}

   \begin{figure}
   \begin{center}
   \begin{tabular}{c}
   \includegraphics[height=7cm]{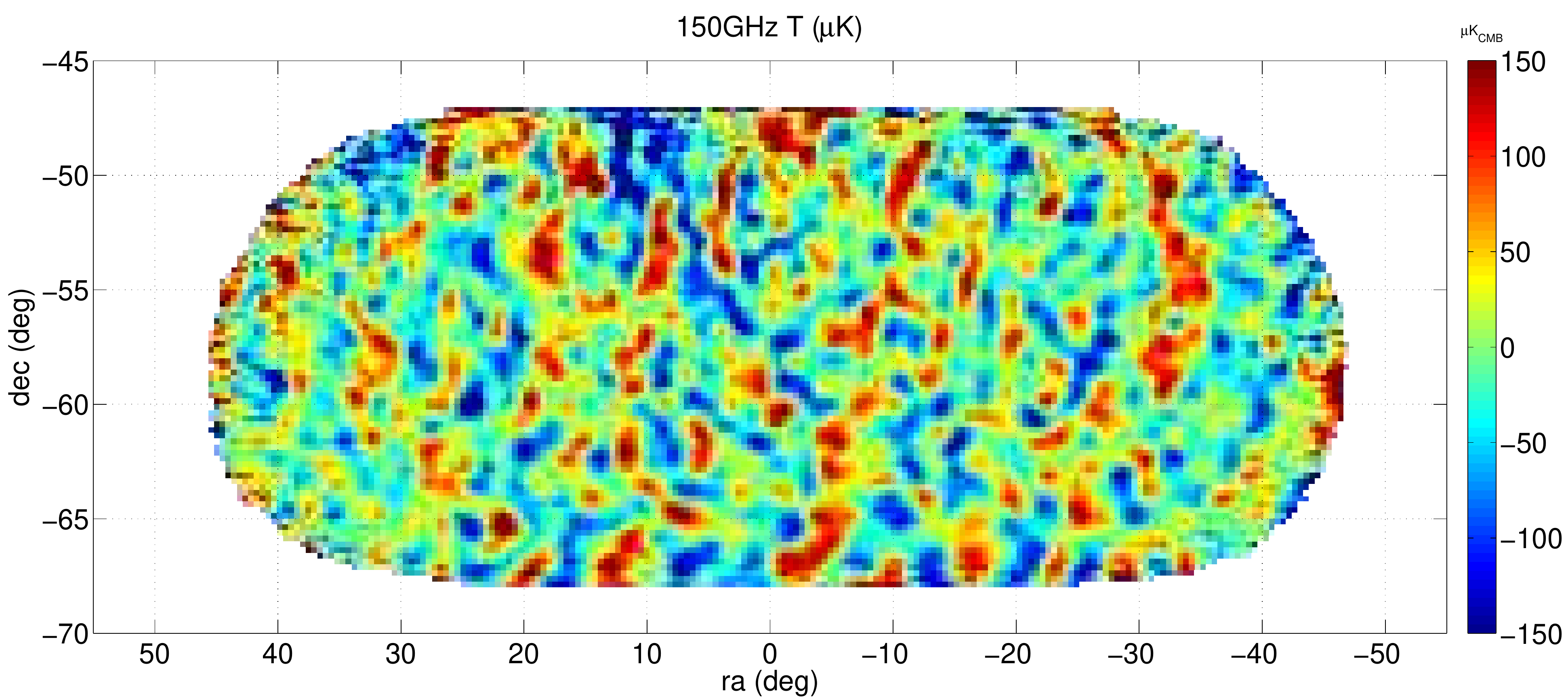}
   \end{tabular}
   \end{center}
   \caption[example]
   { \label{fig:cmb}
CMB temperature map for an 8 day subset of data.}
   \end{figure}

One method for estimating the sensitivity is with a CMB map\cite{2011ltd.brevik}.  Figure \ref{fig:cmb} is an early season 2012 map of the CMB temperature, coadded over all detectors for an 8 day subset at the end of April.  This map will be used for the rest of the section for calculating estimates of NET.  The CMB temperature fluctuations are cross-correlated with WMAP to obtain an absolute calibration\cite{Chiang:2009xsa}.  As was done in Chiang et al.\cite{Chiang:2009xsa}, the WMAP maps (Q, W and V bands) are smoothed with Keck Array's beam, converted to simulated timestreams, processed with Keck Array's filtering and converted back into maps, thus creating ``Keck Array Observed'' WMAP maps.  In order to avoid a noise bias, one of the other WMAP bands is used as a reference.  This is written out in Equation \ref{eqn:abscal}

\begin{equation}
g=\frac{\sum_\ell \langle a_{\ell m}^{\mathrm{WMAP-1}} a_{\ell m}^{Keck} \rangle}{\sum_\ell \langle a_{\ell m}^{\mathrm{WMAP-1}} a_{\ell m}^{\mathrm{WMAP-2}} \rangle}
\label{eqn:abscal}
\end{equation}
where $\ell$ is summed over the range of 30-210 and $g$ is the resulting Keck Array absolute calibration.  The gain is flat over this multipole range.  The results are the same regardless of which Q, W, or V band maps are used.

   \begin{figure}
   \begin{center}
   \begin{tabular}{c}
   \includegraphics[height=7cm]{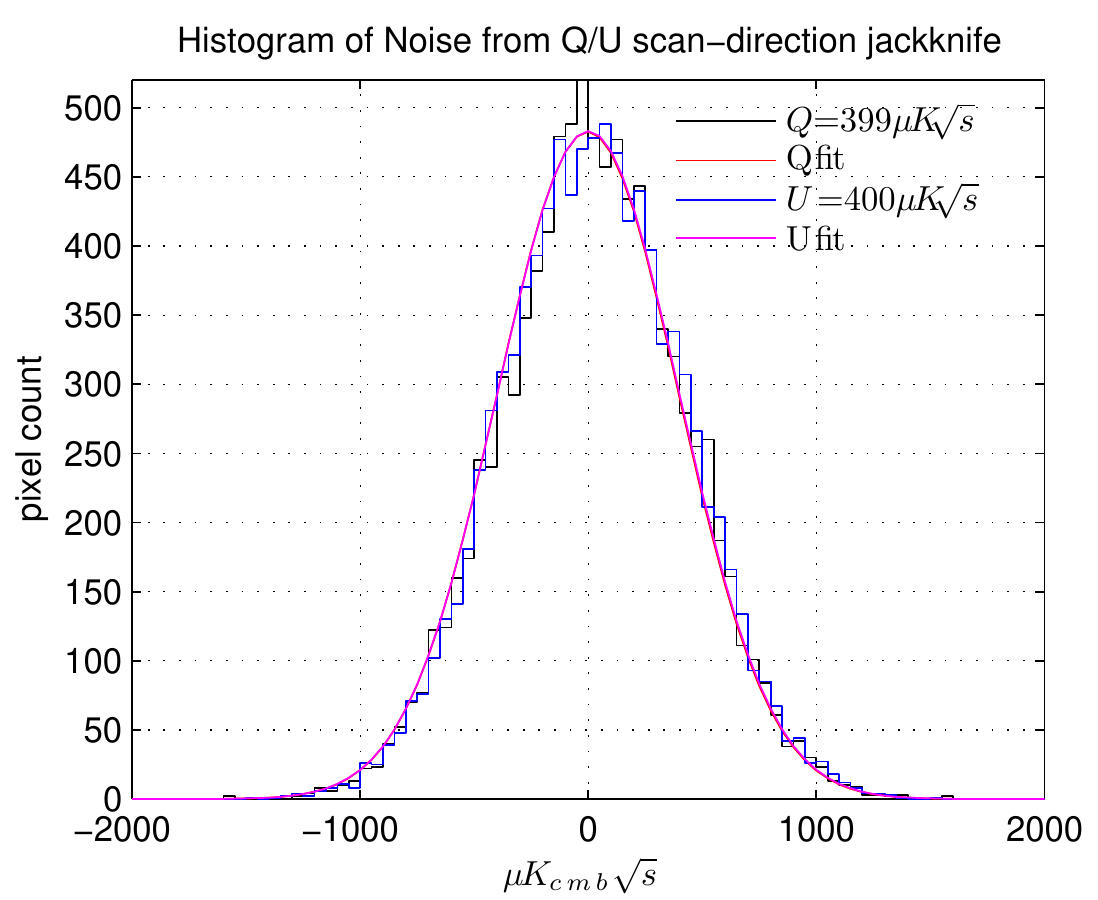}
   \end{tabular}
   \end{center}
   \caption[example]
   { \label{fig:QU}
The histogram of the scan-direction jackknife Q/U map times the square root of the total integration time map.  The integration time is total detector time on each pixel for the entire array, making this is an estimate of the per-detector NET.}
   \end{figure}

After calibrating the maps, the sensitivity of the array can be estimated by directly examining the noise within the maps.  A scan-direction jackknife, the difference map of data taken in the two azimuth scan directions, is used to remove any signal in order to get an estimate of the noise in the maps.  The polarization pair differencing removes most of common mode noise from the atmosphere.  In the T maps, when all of the detectors are summed, the 1/f noise contaminates the science band even in the scan-direction jackknife.  Because measuring polarization is the goal of this experiment, the Q and U maps provide a better estimate of the array NET.

The noise in the scan-direction jackknife Q and U maps multiplied by the square root of the integration time map is defined to be the per-detector NET.  The signal is divided between the Q and U maps, but this cancels out the pair differencing operation.  The histogram of the resulting map is shown in Figure \ref{fig:QU}.  The per-detector NET from this calculation using the 8-day subset of data is approximately 400 $\mu \mathrm{K}_{CMB} \sqrt{s}$.  The number of effective detectors used to make this map is total integration time divided by the integration time for a single detector.  Including the data cuts for this time period gives an overall array sensitivity of 11.5 $\mu \mathrm{K}_{CMB} \sqrt{s}$ in the Q/U maps in the 2012 configuration.  The weather was good during this subset of data, but data is also cut based on other quality measures such as flux jumps (large SQUID steps), noise stability, and relative gain changes.

   \begin{figure}
   \begin{center}
   \begin{tabular}{c}
   \includegraphics[height=7cm]{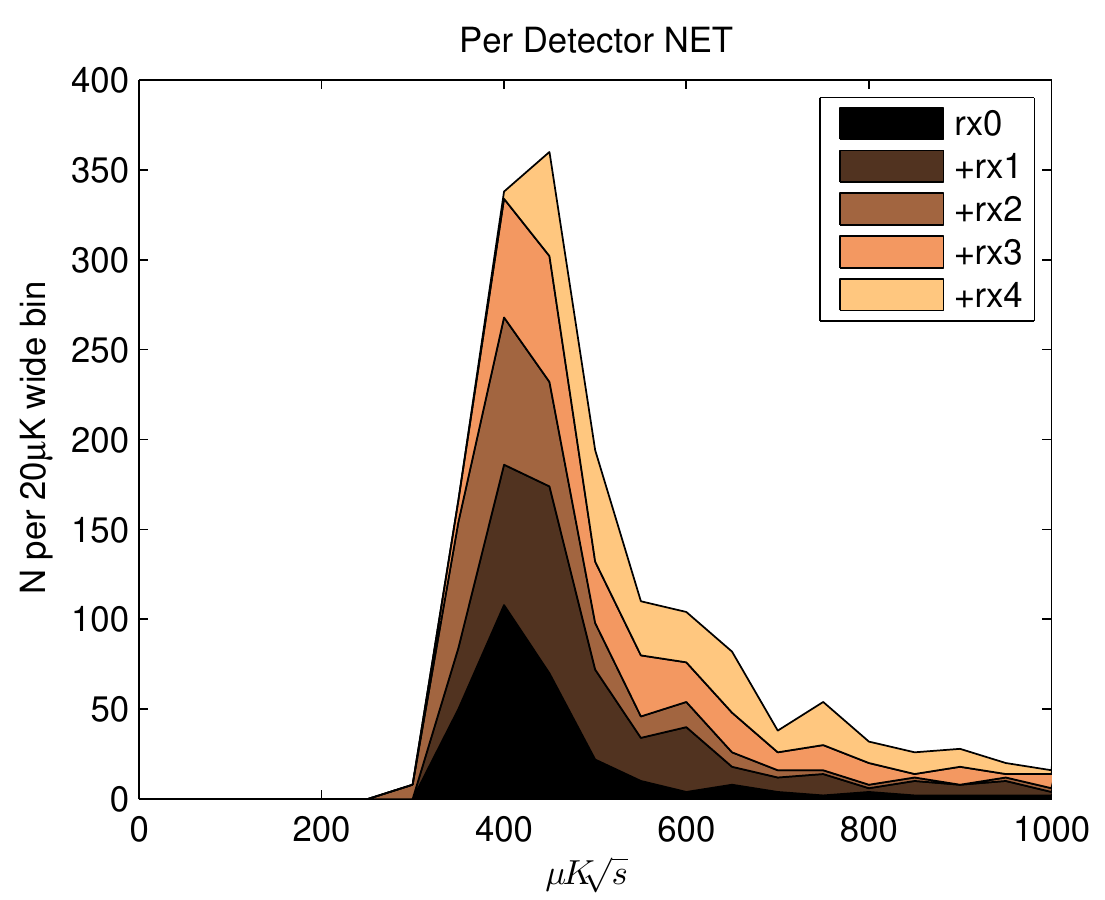}
   \end{tabular}
   \end{center}
   \caption[example]
   { \label{fig:NEThist}
Histogram of the NET per detector using timestream based estimates.  Rx3 and Rx4 have lower optical efficiencies than the other receivers.  We will be upgrading those telescopes with more efficient detectors for 2013.}
   \end{figure}

\subsection{Timestream based estimates}
\label{sec:time}

Another method to estimate the sensitivity is by directly analyzing the noise of the timestream in fourier space\cite{2011ltd.brevik}.  This was also computed on the same 8-day subset of data as the map-based method.  The responsivity of the detectors is estimated using the calibration elnods.  This depends on the sky temperature, which is calculated using the absolute calibrations of our detectors from WMAP as described above.  The noise level of each detector pair is calculated from the polarization pair-differenced timestream in the frequencies of interest (0.1-1 Hz).  Figure \ref{fig:NEThist} is the resulting histogram of sensitivities per detector across the array.  This method results in a per detector NET of approximately 440 $\mu \mathrm{K}_{CMB} \sqrt{s}$, and an array NET of 11.7 $\mu \mathrm{K}_{CMB} \sqrt{s}$ for the 2012 configuration.  The per-detector NET is 10$\%$ higher in this method because of the residual 1/f noise in the data, as seen in Figure \ref{fig:pairdiffnoise}.  However, the overall array sensitivities are comparable, because of the cuts applied while making the maps.

\acknowledgments     

The Keck Array projects have been made possible through support from the National Science Foundation (grant Nos. ANT-1044978/ANT-1110087) and the Keck Foundation.  Detector development has been made possible with the Gordon and Betty Moore foundation.  We also acknowledge the Canada Foundation for Innovation and BC Knowledge Development Fund for support.  We are grateful to Robert Schwarz for spending the winter in the South Pole for us in both 2011 and 2012, as well as to the South Pole logistics team.  We also are grateful for the insight and collaboration from the entire \bicep2, SPIDER and Keck Array teams.


\bibliography{spie2012_sask}   
\bibliographystyle{spiebib}   

\end{document}